\documentclass{emulateapj}

\shorttitle{The GLE linked with M5.1-class solar flare on May 17, 2012}
\shortauthors{Augusto et al.}

\begin{document}


\title{Was the GLE on May 17, 2012 linked with the M5.1-class flare the first in the 24th solar cycle?\\ 
}


\author{C. R. A. Augusto, V. Kopenkin and C. E. Navia\altaffilmark{1}}
\affil{Instituto de Fisica, Universidade Federal Fluminense, 24210-346, Niteroi, Rio de Janeiro, Brazil}

\author{A. C. S. Felicio, F. Freire, A. C. S. Pinto, B. Pimentel, M. Paulista and J. Vianna}
\affil{Departamento de Fisica, Universidade Federal Fluminense, 24210-346, Niteroi, Rio de Janeiro, Brazil}

\author{A. C. Fauth}
\affil{Instituto de Fisica Gleb Wataghin, Universidade Estadual de Campinas, 13083-859, Campinas, S\~{a}o Paulo, Bazil}

\and

\author{T. Sinzi}
\affil{Rikkyo University, Toshima-ku, Tokyo 171, Japan}

\altaffiltext{1}{E-mail address:navia@if.uff.br}


\begin{abstract}
On May 17, 2012 an M5.1-class flare exploded from the sun. 
An O-type coronal mass ejection (CME) was also associated with this flare.
There was an instant increase in proton flux with peak at $\geq 100$ MeV, leading to S2 solar radiation storm level.
In about 20 minutes after the X-ray emission, the solar particles reached the Earth.
It was the source of the first (since December 2006) ground level enhancement (GLE) of the current solar cycle 24.
The GLE was detected by neutron monitors (NM) and other ground based detectors.
Here we present an observation by
the Tupi muon telescopes  (Niteroi, Brazil, $22^{0}.9 S$, $43^{0}.2 W$, 3 m above sea level) 
of the enhancement of muons at ground level associated with this M5.1-class solar flare. 
The Tupi telescopes registered a muon excess over background  $\sim 20\%$ in the 5-min binning time profile.
The Tupi signal is studied in correlation with data obtained by space-borne detectors (GOES, ACE), ground based  neutron monitors (Oulu) and 
air shower detectors (the IceTop surface component of the IceCube neutrino observatory). 
We also report the observation of the muon signal possibly associated with the CME/sheath striking the Earth magnetosphere on May 20, 2012.  
We show that the observed temporal correlation of the 
muon excess observed by the Tupi muon telescopes with solar transient events suggests a real physical connection between them.
Our observation indicates that combination of two factors, the low energy threshold of the Tupi muon telescopes
and the location of the Tupi experiment in the South Atlantic Anomaly region, can be favorable in the study and detection of the 
solar transient events.
Our experiment provides new data complementary to other techniques (space and ground based) in the study of solar physics.
\end{abstract}





\section{Introduction}

Ground-level enhancements (GLEs), typically in the GeV energy range, are sudden increases in cosmic ray intensities 
registered by neutron monitors (NMs), which  are ground based instruments that detect the secondary neutrons produced by the 
primary protons penetrating the Earth's atmosphere \citep{gopalswamy10}.
These enhancements can be registered by other types of ground detectors, 
such as air shower detectors and muon telescopes  \citep{nitta12}. 
In most cases GLEs  take place during the intense X-class\footnote{Flares are classified into X, M, C, B and A flares, 
with X corresponding to GOES flux in excess of $10^4 \, Watts \,m^{-2}$ at Earth, 
and successive classifications decreasing in decades \citep{fletcher2011}.} 
solar flares as well as the fast (above $\sim 1000 \, km \, s^{-1}$) Coronal  Mass Ejections (CMEs) \citep{gopalswamy10,gopalswamy11a}. 
There are some GLEs cases associated with weaker flares and slower CMEs \citep{cliver06} 

GLE events are solar energetic particle (SEP) events. 
SEP are one population of particles in the interplanetary space from the Sun to 1 AU,
that are classified according to particle origin or acceleration region 
\citep{kallenrode2001,kallenrode2003,tbt2009}.
SEP have intermediate energies from $\sim 10 keV/nucleon$ to the GeV range and  occur in events 
that last from some hours to a few days.
SEP are much more frequent at times of solar maximum than during solar minima, the occurrence of SEP events 
is directly related to flares and CMEs \citep{kallenrode2003}.
There are recognized two distinct classes of SEP events, impulsive (accelerated in flares) and gradual (accelerated at 
CME driven shocks) \citep{cane86,kallenrode2003}.

A number of individual GLEs has been studied in details in literature, nevertheless
the exact conditions and processes that are responsible for these extreme SEP events are not understood yet \citep{nitta12}.

Since 1950th the observation and the study of the powerful solar flares
has been done exclusively with NMs all around the world, especially in the polar regions \citep{meyer56,simpson00,moraal00}.
Observation with NMs yield a lot of new information. For instance,
the anti-correlation between solar activity and the flow of galactic cosmic rays,
the existence of a prompt and late emission in flares, correlations of the cosmic ray intensity with CMEs and other 
solar disturbances crossing the Earth, etc. \citep{chupp,moraal00}.
Nowadays, particles accelerated near the Sun can be detected by space-borne instruments.
The  measurement of high energy protons in space is achieved by the High Energy Proton and Alpha
Detector (HEPAD) \citep{onsager96} on the Geostationary Operations Environmental Satellite (GOES), 
which provides data on differential fluxes in three channels in the energy region between 350 MeV and 700 MeV, integral 
flux above 700 MeV, and the  X-ray flux in two wavelengths ($(http://www.oso.noaa.gov/goes/index.htm)$). 

Solar flares that can be measured at the Earth's surface are rare events.
Not all of the solar explosions observed by satellites can be measured at the Earth's surface. 
For instance, there can be  dissipation of the radiation by the interplanetary magnetic field (IMF), 
particles can be  deflected or captured by the Earth's magnetic field or absorbed by the Earth's atmosphere.

GLEs are quite rare events.
Fewer than 100 events have been observed by NMs in the last 70 years ($http://www.nasa.gov/mission_pages/sunearth/news/particles-gle.html$).
It is important to note that ground detectors do not observe GLEs simultaneously. 
For instance, the NMs which observed the GLEs in solar cycle 23 were located at regions with the geomagnetic rigidity cutoff $\sim 1 GV$ or less \citep{shea11}. 

Solar flares and CMEs occur whenever there is a rapid large-scale change in the Sun's magnetic field. 
Solar flare detection at ground level depends on good magnetic connection between the Sun and Earth. 
Most solar flares associated with GLEs are located on the western sector of the
Sun where the IMF is well connected to the Earth \citep{reames99}.
At least two more factors  may
contribute to GLEs: the presence of prior CMEs and the magnetic field connection of the acceleration region to Earth.
The time profiles of the observed SEP events depend on the original solar active region  longitude (heliologitude). 
For example, those from the western source active regions tend to rise more
quickly to the peak than those from the eastern source regions \citep{cane88,cane03,nitta12,reames99}. 
The solar active areas are distributed in much broader longitudes than those of impulsive SEP events in flares \citep{reames99}.

Among 16 GLEs observed in the previous solar cycle 23,  14 GLEs were linked with X-class flare, 
one with an M7.1-class flare and one with a C2.2-class flare. 
It was reported that in 15 GLE the associated blast has also emitted a CME \citep{kahler01}; \citep{cliver83,cliver06}. 
Usually the CMEs provide conditions  for the seed particles to be re-accelerated.
These shock driven CMEs lead to GLEs that can be observed by ground based detectors. 
The last GLE from the previous solar cycle 23 was originated by the X3.4 class flare at (02:14 Universal Time (UT)) on December 13, 2006, 
at the active solar spot region with heliospheric coordinates (S06,W23). 
There was also a CME associated with this X3.4 solar blast. 

On the other hand, 
the first GLE observed in the current solar cycle 24 has been linked with an M5.1-class solar flare on May 17, 2012. 
An O-type (type O stands for "occasional") CME was also launched during this event 
($http://cdaw.gsfc.nasa.gov/$),
and a category S2 solar radiation storm\footnote{The scales have numbered levels ($http://www.swpc.noaa.gov/NOAAscales/$).} was observed. 

Here we present new data on the observation of muon excess in association with the M5.1-class flare. 
In this study we include data obtained by other ground based observations, 
such as Oulu NM \citep{usoskin11}  and the IceTop air shower detectors \citep{tamburro12}, as well as  observations obtained by space-borne detectors. 
In addition, we also analyze the Earth passage of the magnetic disturbance linked with the CME originated in the solar eruption on May 17, 2012.
The Tupi observations are studied in correlation with observations reported by the Advanced Composition Explorer (ACE) located in the L1 
Lagrangian point ($http://www.swpc.noaa.gov$).
We also used the estimated 3 hr planetary $K_p$ index considering that it reflects the mean magnetospheric activity 
($(http://www.swpc.noaa.gov/rt_plots/kp_3d.html)$).
Based on the Tupi data, we obtained the  time delay and the signal significance of the flare and the CME ejecta,
as well as  the earth passage time of the ejecta.

This article is organized as follows. 
In Section 2 we discuss favorable conditions to detect the impulsive SEP, especially those linked with GLEs. 
In addition to those already recognized in literature, 
we highlight the specific location of the Tupi muon telescopes near the South Atlantic Anomaly (SAA) central region.  
Section 3 explains experimental setup. 
Section 4 presents methodology and the results of the association between the Tupi muon excess and an M5.1-class flare on May 17, 2012,
registered by satellites and ground based detectors.
Section 5 describes the Tupi observation of the muon excess on May 20, 2012 possibly associated with the Earth passage of the CME ejecta originated in 
the solar events on May 17, 2012. 
Section 6 presents other examples on the association between muon excesses at ground
level  and transient solar events in the current solar cycle 24, registered by spacecrafts.
In Section 7 we present summary and our conclusions.

\section{Conditions to ``hunt'' for solar energetic particles at Earth} 

Large solar flares are often associated with CMEs.
Under favorable conditions, such as a good magnetic connection between Sun and Earth,
a blast of high-energy particles  with energies above several hundred of MeV, 
can reach the upper atmosphere and produce secondary particles detected by ground based detectors.
The energetic charged particles emitted by a flare,  move into the interplanetary medium, following the solar
magnetic field lines (Parker spiral), which defines the sunward magnetic field direction.
In general, solar flares originated in active areas located in the western sector of the
Sun (in the heliolongitude) have a good magnetic connection with the Earth.
The best condition happens when the emitted particles have very small pitch angles \footnote{The pitch angle is defined here as the
angle between the viewing direction (asymptotic cone of acceptance) and the sunward magnetic field direction.}, 
relative to the foot point spiral field that connects the Earth to longitude  $55^0$ West
of the central meridian of the Sun (as viewed from the Earth) \citep{sheasmart1996,smartshea2003}.
Under this condition the energetic particles  travel essentially along 
the interplanetary magnetic field lines, and their diffusion in the interplanetary medium is negligible.

This minimum scattering condition is usually called a "scatter free" propagation. 
In general, MeV solar particles  appear to have an average mean free path length of the order of $\sim 0.1-0.3$ AU 
and the time delay between the X-ray emission (observed at 1 AU) and the onset of the GLE is several dozens of minutes. 
On the other hand, long time delays indicates a diffusive propagation. 
For solar flares from active regions located at the East of the helilongitude, the time delay (between emission and the GLE onset) can be from several hours up to days. 
Almost all diffusion models involving solar particle transport in the interplanetary medium show that the maximum time delay is proportional to the square of the distance traveled.
Using the heliographic coordinate system \citep{wibberenz74}, the onset time of the GLE is given by \citep{barouch71}; \citep{smart85}
\begin{equation}
T_d (in\;hours) =\frac{0.133\;D}{\beta} + 4\theta^2,
\end{equation}
where $\beta=v/c$, v is the particle speed, $\theta$ is the heliolongitude, and  D is the Archimedean spiral path length (measured in AU). 
The minimum in the curve  corresponds to zero pitch angle condition, that is a flare at the "foot point" of the Archimedean spiral path between the Earth and the Sun (this is about ${\theta}=57.2^{0}$).
Figure 1 shows the time delay as a function of the helilongitude, where the solid line indicates is the expected delay time according 
to Wibberenz's model \citep{wibberenz74}.

\begin{figure}
\hspace*{-1.0cm}
\vspace*{-6.0cm}
\includegraphics[width=4.0in]{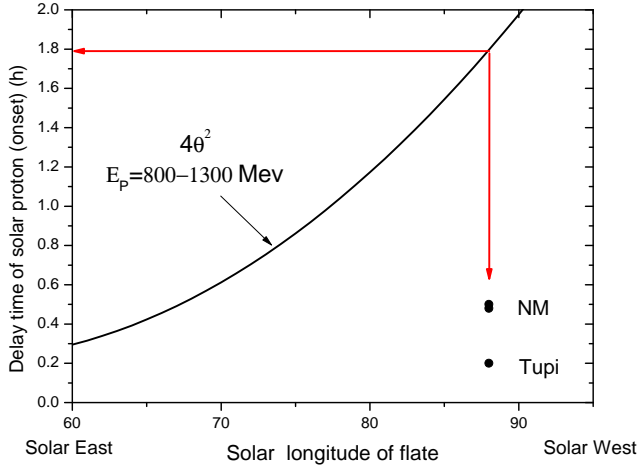} 
\vspace*{-1.0cm}
\caption
{
Expected time delay (onset at Earth) of 800 to 1300 MeV solar protons as 
a function of solar longitude (solid line).
The May 17, 2012 solar flare was associated with a active region AR1476 located at (N07W88), so the expected time delay  is about 1.8 h.
The time delay observed by NMs is $\sim 0.49$ h and the time delay for the Tupi detector is $\sim 0.2$ h.}
\label{fig1}
\end{figure}

On the other hand, both the intensity and the spectrum of SEP observed by ground based detectors 
also depend on the relative locations of these detectors on Earth, as well as 
at the time of the occurrence of solar flares. 
For instance, ground based detectors positioned near the geomagnetic equator are not able to register SEP from solar flares due to high geomagnetic cutoff rigidity. 
Thus, the polar regions with low geomagnetic cutoff rigidity are the most favorable locations to detect solar transient events.

However, there is one more place on Earth, where the Earth magnetic field is weakened considerably. 
This is the South Atlantic Anomaly (SAA), the area at mid-South latitudes, off the coast of Brazil (see Figure 2). 
It covers almost 7.8 million $km^2$ and currently continues to grow.

\begin{figure}
\hspace*{-1.0cm}
\vspace*{-5.0cm}
\includegraphics[width=4.0in]{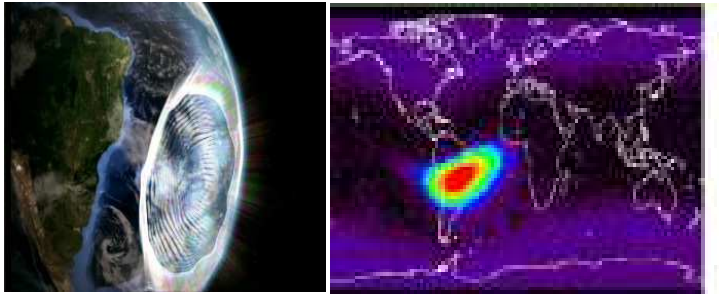} 
\vspace*{-3.0cm}
\caption
{
Left panel: Sketch of the geomagnetic hole in the South Atlantic area. 
The magnetic field intensity in this area is about three times lower  than the intensity of the magnetic field at the SAA boundary. 
Right panel: Geographical distribution (latitude vs longitude) of proton flux ($E>850MeV$) measured by the HEPAD ICARE detector\cite{boatella10}. 
The color scale is logarithmic. 
The dark color(red) represents 10 times higher proton flux than the one with the light color(blue).
}
\label{fig2}
\end{figure}

Geographical distribution of proton flux measured by the HEPAD ICARE instrument on-board the Argentinean satellite SAC-C showed an excess (up to 10 times) of protons with $E>850$ MeV in the SAA central region in comparison with the region outside of the SAA. 
These high energy protons are hard to be considered as Van Allen trapped protons, because the SAA models such as AP8 and several measurements of the trapped protons showed that their energies do not exceed 300 MeV \citep{boatella10}.
In addition, the PAMELA Collaboration introduced a subcutoff in the rigidity that is
below the nominal Stormer rigidity cutoff in the SAA area \citep{casolino}.
 Downward going protons with energies above 200 MeV are also clearly seen at the latitudes between $40^{0}$S and $20^{0}$ S (the SAA area). 
 These are so-called "quasi-trapped particles".
 This geomagnetic subcutoff is about 10 times smaller than the nominal geomagnetic cutoff rigidity. 
 At the Tupi experiment location ($22^{0} .9S$, $43^{0} .2 W$) that is close to the SAA central region, this subcutoff is around 800-1000 MV (equivalent to 800-1000 MeV for protons.)

Thus, the primary  and secondary charged cosmic ray particles 
can penetrate deep into the atmosphere owing to the low geomagnetic field intensity over the SAA. 
Consequently, in the SAA region cosmic ray fluxes at lower energies  are even higher than world averages at comparable altitudes. 
This is reflected as an enhancement in the counting rate of the incoming primary cosmic ray flux.

\section{Experimental setup}

The Tupi experiment  has two muon telescopes \citep{augusto11}. 
Each telescope was constructed on the basis of two detectors (plastic scintillators 50 cm x 50 cm x 3 cm) separated by a distance of $3 \,$m. 
One telescope has a vertical orientation, and the other one is oriented near 45 degrees to the vertical (zenith), pointing to the west. 
Each telescope counts the number of coincident signals in the upper and lower detector. 
The output raw data consists of counting rate versus universal time (UT). 
The Tupi telescopes are placed inside a building, under two flagstones of concrete ($150 g/cm^{2}$).
The flagstones increases the detection muon energy 
threshold up to $\sim 0.2-0.3$ GeV required to penetrate
the two flagstones.
The Tupi telescopes has an effective field of view $0.37 sr$. 
The effective solid angle of each detector  
can be roughly obtained from the following relation: $\Omega=2\pi(1-\cos \theta_z)$, where $\theta_z$ is the maximum zenith angle.

Time synchronization is essential for correlating event data in the Tupi experiment, and this is achieved by using the GPS receiver. 
All steps from
signal discrimination to the coincidence and anticoincidence
are made via software, using the virtual instrument
technique. 
The application programs were written using the LAB-VIEW tools. 
The Tupi experiment has a fully independent power supply, with an autonomy of up to 3 h to
safeguard against local power failures.
As a result, the data acquisition is basically carried out 24 h a day.

The Tupi experiment is in the process of constant expansion and upgrade. 
There is work in progress on additional new telescopes in Campinas (Brazil) and La Paz (Bolivia).

\section{Observation of the GLE on May 17, 2012 by the Tupi muon telescope}

Here, we present the muon excess associated with the  May 17, 2012 M5.1-class solar  flare  observed by the Tupi muon telescopes.
We  compare our data  with observations made by spacecraft detectors, such as GOES satellites. 
The timeline of the solar flare (onset;peak;end) is as follows: (01:25 UT;01:47 UT;02:14 UT) ($http://www.lmsal.com/solarsoft/latest_events_archive.html$).
The X-ray energy flow  measured by the GOES in the wave length  0.8-4 A is used to classify the solar flare.
For the May 17, 2012 event the X-ray energy flow peak went up to
$5.1 \times 10^{-5} \, Watts \, m^{-2}$, and represents an M5.1-class flare. 
In general, M-class flares are ten times smaller (in X-ray flow) than the X-class flares
(the most powerful). 
The May 17, 2012 solar eruption consisted of an M5.1 soft X-ray solar flare and a fast CME.
In addition, the GOES registered increase in the proton flux. 
There was observed a S2 radiation storm.
Figure 3 summarizes the situation.

\begin{figure}
\hspace*{+1.0cm}
\vspace*{-3.0cm}
\includegraphics[width=3.0in]{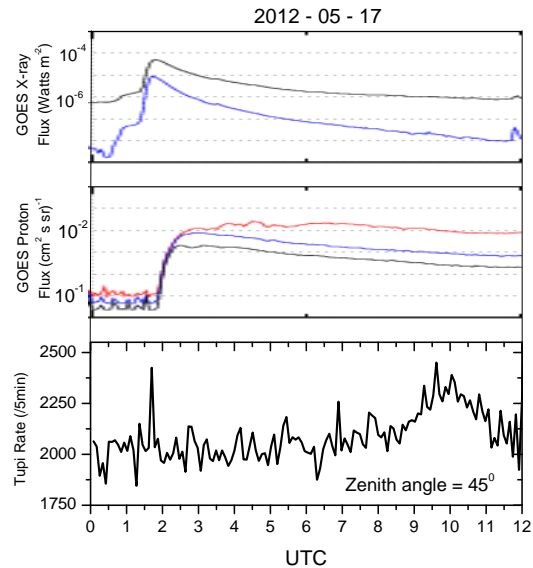} 
\vspace*{+2.0cm}
\caption
{
Observation of the  May 17, 2012 solar flare.
The x-ray prompt emission of the solar flare at the onset time
(01:25 UT) is classified as M5.1 class by the GOES satellite (top
panel). 
Proton flux time profile observed by the GOES  satellite (central panel).
For comparison, there is time profile of the 5-min muon counting rate observed by the inclined
Tupi muon telescope (bottom panel). 
}
\label{fig3}
\end{figure}

The inclined Tupi telescope registered a muon excess  over background $\sim 20\%$ in the 5-min binning  time profile. 
We estimate signal significance in the period ($0-9 \, h$ UT).
The percent deviation of a signal $S$ is defined here as $S={(C^{(i)}-B)}/{{B}}$, where $C^{(i)}$ is the measured number of counts in the bin "i" and B is the average background count for a given time period. 
 The occurrence of GLE (muons excess) as observed by the Tupi telescope can be interpreted as evidence of the
arrival in the upper layers of the atmosphere of a bundle of protons and/or ions with energies exceeding the pion production
and above the local geomagnetic cutoff rigidity.
The onset time observed by the inclined Tupi telescope is about $\sim 1.62 \, h$  UT, an the peak time at $\sim 1.7 \, h$ UT.
In Figure 3 one can notice an increase in the counting rate after 9 h UT. 
This seasonal variation of the intensity is called day-night asymmetry \citep{augusto10}.
The increase in precipitation usually happens about 3 h after sunrise, and similar effect (the decline in precipitation) occurs about 1 h after sunset.

On May 17, 2012, there have been observed a significant increase in the counting rate of the IceTop air shower detector\footnote{The IceTop surface component of the IceCube neutrino observatory.}, a neutron  monitors at the South Pole NM and Oulu NM (close to the North Pole)
in association with the M5.1-class flare. 
This observation constitute the first GLE in the current solar cycle 24.

A comparison between the GLE observation made by the IceTop air shower detector at the South Pole and the muon counting rate in the
inclined Tupi telescope is shown in Figure 4. 
The IceTop has two discriminators rates: SPE (Single Photo Electron) and MPE (Multi Photo Electron).
The counting rates (for both SPE and MPE) are in the range from 1 to 10 kHz. 
On the other hand, the  counting rate of the Tupi telescopes is 100 kHz. 
In Figure 4 we show the time profiles of the vertical and
inclined Tupi telescopes on May 17, 2012.
The muon excess  over background in the vertical Tupi telescope is $\sim 10\%$, and in the inclined Tupi telescope - $20\%$.

\begin{figure}
\hspace*{+0.5cm}
\vspace*{-0.0cm}
\includegraphics[width=3.0in]{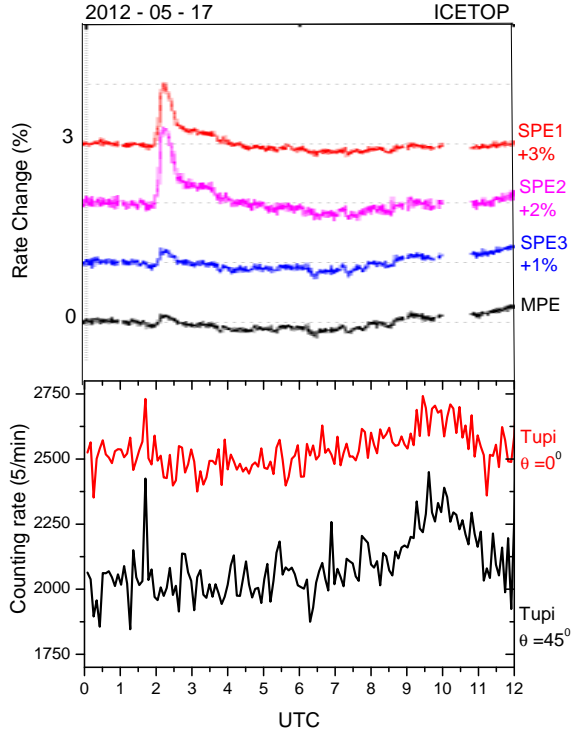} 
\vspace*{-0.0cm}
\caption{Top panel: 
time profile of the SPE-MPE counting rate change (in percent) observed by the IceTop air shower detector.
Bottom panel: time profiles of the 5-min muon counting rate observed in the vertical and inclined Tupi telescopes on May 17, 2012.. 
All observations were made on May 17, 2012.
}
\label{fig4}
\end{figure}

The GLE event observed by the Oulu NM (located in the region near the North Pole) on May 17, 2012 is presented in Figure 5.
For comparison we show the time profile of the 5-min muon counting rate in the
inclined Tupi telescopes on May 17, 2012.

These ground based observations show large extension and anisotropy of the SEP linked with the M5.1 solar flare. 
Usually the counting rate of secondary particles produced by vertically incident  projectiles 
is in reverse correlation with the atmospheric pressure. 
The variation of the barometric coefficient   observed by  NM detectors is around $1\%$ per mbar. 
However, the variation of the barometric coefficient for muons at sea level  is $\sim 10$ times lower, i.e. $0.1\%-0.2\%$ per mbar.

\begin{figure}
\hspace*{+0.5cm}
\vspace*{-0.0cm}
\includegraphics[width=3.0in]{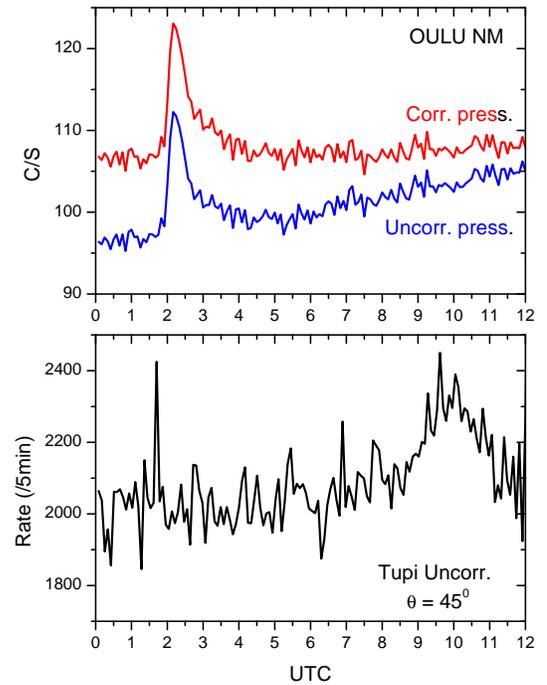} 
\vspace*{-0.0cm}
\caption{Top panel:
The May 17, 2012 GLE event observed by the Oulu NM (corrected and uncorrected for pressure). 
The onset time is 01:43 UT (when the intensity is $2\%$ of the peak).
Bottom panel:  time profile of the 5-min muon counting rate observed by the inclined
Tupi muon telescope.  
}
\label{fig5}
\end{figure}

We would like to point out some differences between the Tupi observations and the IceTop-Oulu observations.
The Tupi telescopes observe the muon excess as a narrow peak.
As one can see from Figures 4 and 5, the width of the GLE signal observed by IceTop-Oulu is $\sim 3$ times wider that the Tupi peak. 
This difference can be attributed to the small field of view of the Tupi telescopes.
The effective field of view of NMs and air shower detectors is  higher, $\sim 3.14 \, sr$. 
The muon telescope detects only those particles whose direction of propagation  is close to the axis of the telescope,
by other words, when the particle flux is contained within the telescope field of view. 

Comparison of the  May 17, 2012 GLE signal significance as a function of the onset time and the peak time is summarized in Figure 6. 
From this figure we can see that Tupi  observed the peak of the signal before the solar flare maximum (detected by 
GOES), and the IceTop and Oulu experiments observe the GLE peak after the solar flare maximum.
For 1.2 AU Parker spiral length, the solar particle release time
has been reported as  (01:40 UT), normalized to electromagnetic emissions \citep{gopalswamy2012}.

\begin{figure}
\hspace*{-1.0cm}
\vspace*{-0.0cm}
\includegraphics[width=4.0in]{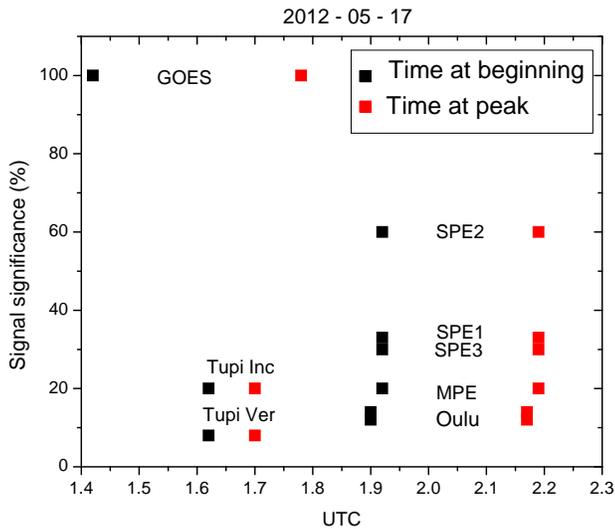}
\vspace*{-6.0cm}
\caption
{
Correlation between the onset time and the peak  time for the  May 17, 2012 GLE observed by different experiments.
The X-ray prompt emission of the solar flare  at the onset time (01:25 UT) is classified as an M5.1 class by the GOES satellite.
The solar flare peak time determined by GOES is (01:47 UT).
The Tupi signal significance in the period ($0-9 h$ UT)  is estimated using the 5-min binning  time profile.
The Oulu NM onset time was reported as ($01:43$ UT) (when the intensity is $2\%$ of the peak).
We made estimations of the signal significance for the IceTop experiment using the  SPE-MPE counting rate in Figure 4.
}
\label{fig6}
\end{figure}

\section{Tupi observation of the O-type CME on May 20, 2012}

Recently the Tupi experiment has reported the results of an ongoing survey on the association between the muon flux variation at ground level
 and the Earth-directed transient disturbances in the interplanetary medium propagating from the Sun 
(such as CME and corotating interaction regions (CIRs)) \citep{augusto12b}.
In present work we apply the same method for the analysis of the geomagnetic disturbance linked with the CME associated with an M5.1
solar flare on May 17, 2012.
According to the SOHO LASCO/C2 ($http://sohowww.nascom.nasa.gov/$), the CME first appearance was at (01:48 UT). 

A sudden increase in the density, speed, and
temperature of the solar wind and in the IMF intensity registered by a satellite, indicates
the arrival of a shock wave .
If the origin of the disturbance is the CME ejecta, 
then the disturbance crosses  the Lagrange point L1 first,  and after several hours it arrives at Earth.
If CME has significant southward component (Bs)  of the interplanetary magnetic field (IMF) 
in either the sheath behind the shock or in the driver gas (magnetic cloud), 
then after reaching the Earth's magnetosphere, it may lead to geomagnetic storms (\cite{tbt1988,wdg1994,tbt1997a,tbt2006a,tbt2006b}).
In addition , the ground-based
detectors can observe depressions of cosmic-ray intensity (so called Forbush events) \citep{cane00,cane96} 
due to the shadow effect by the passage of the disturbance on the Earth.   
Several indices were developed to reflect global information
about current magnetospheric activity based on different
inputs at different locations around the globe. 
Kp and Dst are the most widely used indicators \citep{bartels1963,sugiura1964,rostoker1972,mayaud1980,wdg1994}.

According to the Stereo A  spacecraft, the speed of the O-type  CME on May  17, 2012 was $\sim 1500 km \,s^{-1}$.
The preliminary analysis indicated  that the CME was not entirely directed to the Earth.
However, according to the SWEPAM solar wind detector on board the ACE spacecraft, on May 20, 2012 at (01:44 UT) there was registered an interplanetary shock. 
A possible origin of this shock was the O-type CME of May 17, 2012. 
The shock signal at Earth was very soft, with the Kp index $K_p= 4$ at (3-6 h UT). 
This means that the disturbances in the geomagnetic field caused by the CME eject was of minor scale.

\begin{figure}
\hspace*{+0.5cm}
\vspace*{-0.0cm}
\includegraphics[width=3.0in]{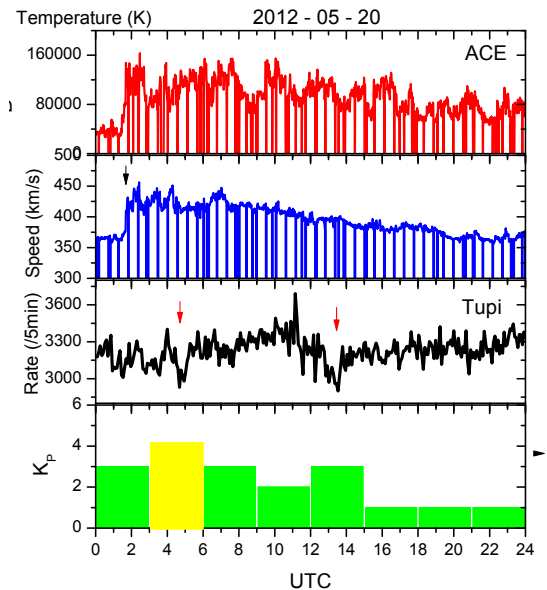}  
\vspace*{-2.0cm}
\caption{
Comparison between 
the solar wind parameters (temperature and speed) 
 observed by the ACE satellite (two panels from the top), 
the $K_p$ index (bottom panel) on May 20, 2012 and the muon counting rate in the vertical Tupi telescope (third panel).
The Tupi counting rate is 5 min binning here.
The arrows in the third panel indicates the begin and the end of the disturbance signal on Tupi data.
}
\label{fig7}
\end{figure}

The signal (with significance $4.1 \%$) in the vertical Tupi telescope is observed at $(\sim 4 \, h UT)$ in association 
with a  $K_p=4$ (see the first arrow in Figure 7).
The second arrow at $\sim 13 h$ UT indicates $\sim 10\%$ increase in the Tupi muon couning rate, the  corresponding $K_p=3$ at (12-15 h UT).
Possible explanation of this observation is that the travel duration of the magnetic disturbance through the Earth  was around 9 hours.
After the passage, the $K_p$ index falls to $K_p=1$.  
In Figure 7 we show comparison between 
the solar wind parameters (temperature and speed) 
 observed by the SWEPAM solar wind detector on board the ACE satellite, 
the $K_p$ index and the muon counting rate in the vertical Tupi telescope  on May 20, 2012.
 
\section{Other observations of muon excess in association with solar flares in the current solar cycle 24}

Solar flares are characterized by their brightness in x-ray flux. 
In most cases, only flares of X-class and M-class (flux above $10^{-4}$ and $10^{-5} Watts \, m^{-2}$ respectively) are observed by ground
detectors.
Since 2005 (\cite{augusto05,navia05}) the experiment with the Tupi muon telescopes  have reported observation of signals from small solar transient events, including  association with gamma-ray  emission by C-class flares observed by the Fermi GBM detector \citep{augusto11},
as well as  simultaneous observations between Tupi muon excess and the Pierre Auger observatory (SD in scaler mode) counting rate enhancements in association with X-ray emission from solar flares detected by GOES satellites \citep{augusto12}.
It means that small-scale solar flares, those with prompt X-ray emission 
in excess of $10^{-6} \, Watts \, m^{-2}$ at 1 AU (classified as C class)  may give rise to GLEs, 
probably associated with solar protons and ions arriving to the Earth as a coherent particle pulse.
In general, significance of the signal from a solar flare observed by the Tupi telescopes depend on several factors.
For instance,
solar flares originated in active areas with a good magnetic connection with the Earth, during the time interval ($\sim 12-22 \, h \, UT$), 
with practically nondiffusive transport of the particles emitted by he Sun, are favorable for the Tupi experiment observation.
In Figures 8,9,10 we show new examples on possible association between muon enhancements
at ground observed by the TUPI telescopes and the Sun's X-ray activity, a X2.1-class flare on September 6, 2011, a M6.9-class flare on July 8, 2012 and a C2.3-class flare on November 5, 2012 respectively.
Consequently, all observed in the current 24-th solar cycle.
Currently the Tupi experiment is focused on systematic collection of the solar flare events.
%

It is usually assumed that a solar flare must be very powerful, in order to be a candidate for an association with a GLE. 
However, there are  at least three factors favorable for  high sensitivity attained by the Tupi telescopes to solar transient events. 
First, this is the physical location of the Tupi experiment
within the SAA region, where the
shielding effect of the magnetosphere has "dip" due to
the anomalously weak geomagnetic field strength.
Despite the Stormer geomagnetic rigidity cutoff  of 8-9 GV at the Tupi location, 
the SAA introduces a subcutoff in the geomagnetic rigidity  with a low value $\sim 0.8 - 1.0$ GV. 
This means that the SAA works as a "`funnel"' for SEP . 
The second factor is the low energy threshold ($E_{th} \sim 0.2-0.3$ GeV) in the Tupi experiment.
The minimum proton energy $E_p$ needed to produce muons
of energy $E_{\mu}$ in the atmosphere is $E_{p} \sim 10 \times E_{\mu}$. 
Some experiments on  study of solar modulation of galactic cosmic rays use  lead plates of up to 5 cm of thickness, 
that significantly increase the energy threshold (up to several dozens GeV). 
For the purpose, the Tupi experiment does not use the lead plates.
Third, the  Tupi data acquisition  is made on the basis of a Universal Serial Bus card, with a
counting rate of up to 100 kHz per channel.
Since October 2012 the sampled rate of the vertical and inclined telescopes works at 30 Mhz.  

\begin{figure}
\hspace*{+0.5cm}
\vspace*{-0.0cm}
\includegraphics[width=3.0in]{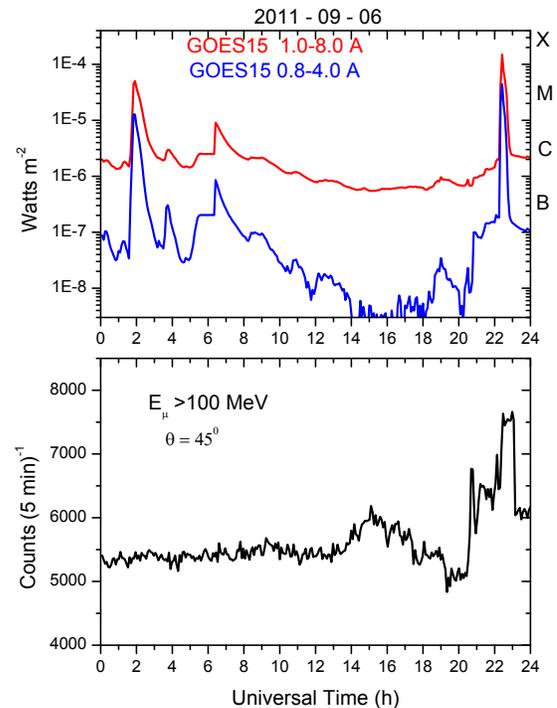} 
\vspace*{-1.0cm}
\caption
{
Top panel: Time profile of the x-ray flux on September 6, 2011, according to GOES 15, for two wavelengths. 
Central and bottom panel: The 5-min muon counting rate in the vertical and inclined Tupi telescopes.
There was an X2.1 solar flare from active area 1283 (N13W18). 
The flare timeline (onset;peak;end): (22:12 UT; 22:16 UT; 22:24 UT).
}
\label{fig8}
\end{figure}

\begin{figure}
\hspace*{+0.5cm}
\vspace*{-0.0cm}
\includegraphics[width=3.0in]{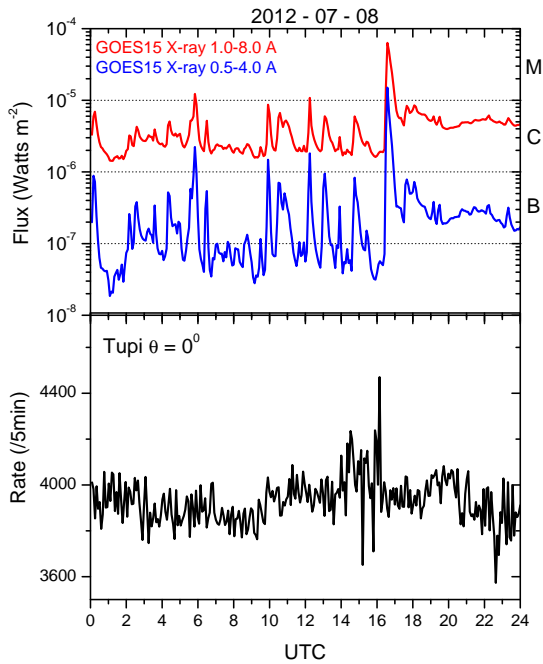} 
\vspace*{-1.0cm}
\caption
{
Top panel: Time profile of the x-ray flux on July  8, 2011, according to GOES 15, for two wavelengths. 
Central and bottom panel: The 5-min muon counting rate in the vertical  Tupi telescope.
There was an M6.9 solar flare from active area 1515 (S14W86). 
The flare timeline (onset;peak;end): (16:23 UT; 16:32 UT;16:42 UT).
}
\label{fig9}
\end{figure}

\begin{figure}
\hspace*{+0.5cm}
\vspace*{-5.0cm}
\includegraphics[width=3.0in]{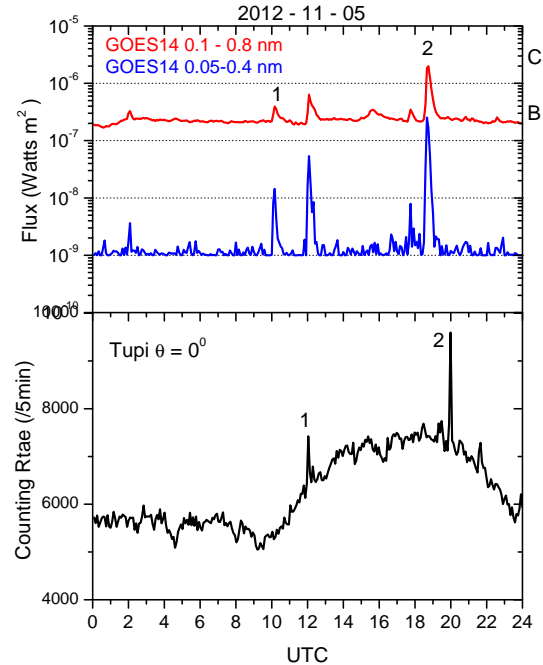} 
\vspace*{+4.0cm}
\caption
{
Top panel: Time profile of the x-ray flux on November 5, 2012, according to GOES 15, for two wavelengths. 
Central and bottom panel: The 5-min muon counting rate in the vertical Tupi telescope.
he numbers 1 and 2 correspond to the 1st and second peaks in association with the GOES data.
Peak 1: a B4.0 solar flare from active area (S14E88). 
The flare timeline(onset;peak;end): (10:00 UT;v10:06 UT;v10:15 UT).
Peak 2: a C2.3 solar flare from active area (S14E88). 
The flare timeline (onset;peak;end): (18:30  UT; 18:39 UT; 18:47 UT).
There is an increase in the counting rate after ($\sim 10 \, h \, UT$). 
This seasonal variation of the intensity is due to day-night asymmetry \citep{augusto10}.
}
\label{fig10}
\end{figure}

\section{conclusions}

Here we described  an observation by the Tupi muon telescopes  
of the muon excess at ground level associated with  an M5.1 class solar flare on May 17, 2012.
This means that the particles, producing muons in the Earth's atmosphere, were emitted by the Sun, and their
transport to Earth was practically nondiffusive. 
This  first GLE event of solar cycle 24 was registered by neutron monitors all over the world.

We have reported a description and an analysis of the
time evolution of the muon flux variation at ground level
in temporal correlation with the May 20, 2012 CME linked with the M5.1 solar flare. 

The Tupi muon telescopes are sensitive to primary
particles (including photons) with energies above the pion
production threshold.
The Tupi telescope can detect muons at sea level with energies greater than $\sim 0.2-0.3$ GeV.
The Tupi telescopes are located at sea level and within the SAA region, where the
shielding effect of the magnetosphere has a "dip" due to the anomalously weak geomagnetic field strength.
This characteristic allows the observation of transient events of diverse origins. 
In this area the primary  and secondary charged cosmic ray particles can penetrate deep into the atmosphere owing to the low magnetic field intensity over the SAA. 
In this region the solar flare induced particles will come down to very low altitudes.
Consequently, in the SAA region cosmic ray fluxes at lower energies  are higher than world averages at comparable altitudes. 
This feature is reflected as an enhancement in the counting rate of the incoming primary cosmic rays flux.

We have also shown at least three experimental evidence of the association between the different types of solar flares and GLEs detected by the Tupi telescopes, already in the current 24th solar cycle.
Our experimental data show that the Tupi
experiment is able to add new information and can be complementary to other techniques designed to study solar transient events.
It is obvious that further experimental observations of experimental events in association with solar transient events are necessary to get better statistical constraints in the study of space weather conditions in Earth's environment. 

\acknowledgments

This work is supported by the National Council for Research (CNPq)
 of Brazil, under Grant  $306605/2009-0$ and $01300.077189/2008-6$,  
Fundacao de Amparo a Pesquisa do Estado do Rio de Janeiro (FAPERJ) under grants $08458.009577/2011-81$ and  $E-26/101.649/2011$
and Fundação de Amparo a Pesquisa do Estado do Estado de São Paulo(FAPESP, grant 2011/50193-4). 
We  express our gratitude to the NOAA Space Weather Prediction Center,
the Lockheed Martin Solar and Astrophysics Laboratory ($http://www.lmsal.com/solarsoft/latest_events_archive.html$),
the STEREO/SECCHI team (Goddard Space Flight Center, Naval Research Laboratory), 
the SOHO/LASCO team (Naval Research Laboratory, Max Planck Institute for Solar System Research), 
the ACE/MAG team, the ACE Science Center, the NMs teams of the Oulu (Sodankyla Geophysical Observatory; $http://cosmicrays.oulu.fi$) and  
Bartol Research Institute ($http://www.bartol.udel.edu/gp/neutronm/$)  for valuable information and data used in this study.


\end{document}